\documentclass{article}
\usepackage{arxiv}
\usepackage[utf8]{inputenc}
\usepackage[english]{babel}
\usepackage[T1]{fontenc}

\usepackage{amsmath}
\usepackage{hyperref}
\usepackage{braket}
\usepackage{quantikz}
\usepackage{lipsum}
\usepackage[numbers]{natbib}

\begin{document}

\title{Deterministic Quantum Search via Recursive Oracle Expansion}

\author{
 John Burke \\
  School of Computer Science and Statistics, \\Trinity College Dublin,\\College Green, \\Dublin 2, \\Ireland\\
  \texttt{burkej15@tcd.ie} \\
   \And
 Ciaran Mc Goldrick \\
  School of Computer Science and Statistics, \\Trinity College Dublin,\\College Green, \\Dublin 2, \\Ireland\\
  \texttt{ciaran.mcgoldrick@tcd.ie } \\
}
\maketitle

\begin{abstract}
We introduce a novel deterministic quantum search algorithm that provides a practical alternative to conventional probabilistic search approaches. Our scheme eliminates the inherent uncertainty of quantum search without relying on arbitrary phase rotations, a key limitation of other deterministic methods. The algorithm achieves certainty by recursively expanding the base oracle so that it marks all states prefixed by the same two bits as the target, encompassing exactly one-quarter of the search space. This enables a step-by-step reduction of the superposition until the target state can be measured with certainty. The algorithm achieves deterministic success with a query complexity of $O(N^{\log_2(3)/2}) \approx O(N^{0.7925})$, falling between Grover’s $O(\sqrt{N})$ scaling and the classical $O(N)$. Our approach relies exclusively on two-qubit nearest-neighbour diffusion operators, avoiding global diffusion entirely. We show that, despite the increased query complexity, this design reduces the total number of two-qubit gates required for diffusion by more than an order of magnitude for search spaces up to at least 18 qubits, with even greater advantages on hardware with limited qubit connectivity. The scheme’s inherent determinism, reliance on simple nearest-neighbour, low-depth operations, and scalable recursive structure make it well-suited for hardware implementation. Additionally, we show that the algorithm naturally supports partial database search, enabling deterministic identification of selected target bits without requiring a full search, further broadening its applicability.
\end{abstract}

\section{Introduction}
Quantum search is one of the most attractive and impactful uses of quantum computing. Grover's algorithm \cite{Grover} established that quantum computers can search an unstructured database in $O(\sqrt{N})$ time, offering a quadratic speedup over the classical $O(N)$ approach. Grover's pivotal approach was later generalised into the broader framework of amplitude amplification \cite{AmplitudeAmplification}, and was proven to achieve optimal quantum query complexity for unstructured search, up to a constant factor \cite{Optimal}. Although this quadratic speedup is more modest compared to the exponential advantage of Shor’s algorithm for integer factorization~\cite{Shor}, quantum search can be applied to a far broader range of problems, including classification~\cite{groverImage}, machine learning~\cite{groverMachineLearning}, and NP computational problems~\cite{GroverNP}.

In general, quantum search algorithms consist of two key components: a quantum oracle, a black-box subroutine that identifies and marks the target item, and a diffusion operator that amplifies the amplitude of the marked state within a superposition. For a search space of size $N$, Grover's algorithm requires $O(\sqrt{N})$ oracle and diffusion iterations to maximise the probability of observing the target state, a substantial improvement over the $O(N)$ steps required to identify the target classically. Furthermore, when multiple marked states ($M$) are involved, the complexity reduces to $O(\sqrt{N/M})$ \cite{GroverMultiple}.

However, a fundamental limitation of Grover’s algorithm is its probabilistic nature. After approximately $\sqrt{N}$ iterations, the success probability is maximised but not guaranteed, and further iterations actually reduce the chance of success, a phenomenon known as "overcooking"~\cite{FixedPointSearch}. Although the probability of error decreases with the size of the search space~\cite{GroverMultiple}, it remains a significant concern for smaller systems or applications where reliability and predictability is critical, such as cryptography~\cite{cryptography1,cryptography2}. 

To mitigate this, deterministic variants of quantum search have been proposed, typically achieving certainty by modifying the oracle and diffusion operators. Some approaches replace the standard phase flip with carefully tuned phase rotations~\cite{AbitraryPhases, GroverZeroFailRate, DeterministicRestrictedOracle, DeterministicAdjustableParameters}, attaining determinism in theory but requiring high-precision, general-purpose quantum gates, well beyond the capabilities of current hardware. Other methods combine both global and local Grover iterations to achieve determinism~\cite{NearDeterminismLocal, LocalDeterminism2}. While this avoids the need for fine-grained phase control, it introduces significant computational overhead in determining the precise sequence and configuration of operations, which quickly becomes infeasible as the number of qubits increases, limiting its scalability and practical use.

In this work, we present a novel quantum search algorithm that can determine a target item with certainty without requiring expensive arbitrary phase rotations. Our approach recursively expands the base oracle such that it marks exactly one-quarter of the current search space at each step, enabling the deterministic reduction of the superposition using only the local two-qubit diffusion operator, avoiding global diffusion entirely. We also show that the algorithm performs partial database search as a natural interim step, enabling deterministic identification of select target bits and further broadening its practical applicability.

Our approach achieves a query complexity of $O(N^{\log_2(3)/2}) \approx O(N^{0.7925})$, outperforming the classical $O(N)$ scaling, though asymptotically higher than Grover’s $O(N^{0.5})$. We show that, despite the higher query complexity, it is significantly easier to implement the two-qubit nearest-neighbour diffusion operators on quantum hardware, than it is to implement global diffusers, requiring over an order of magnitude fewer two-qubit gates to implement the combined diffusion steps on search spaces up to at least $18$ qubits. The approach highlights a practical consideration: while our approach sacrifices some asymptotic efficiency compared to Grover’s algorithm, it achieves a far simpler construction for the diffusion step, relying only on simple, nearest-neighbour operations that are straightforward to implement on quantum hardware. This simplicity, combined with a scalable recursive structure, positions our algorithm as a leading candidate for integration into larger quantum algorithms and applications.

\section{Grover's Algorithm}
To provide context for our approach, we begin by revisiting Grover’s original algorithm. Consider the problem of searching for a particular element $x$ within an unstructured database of size $N$. We define a black-box boolean function $f(i)$, which verifies the solution, as:
\begin{align}
    f(i) &= \begin{cases}
    1 & i = x\\
    0 & \text{otherwise}
    \end{cases}
\end{align}
Classically, identifying the target element $x$ requires checking each database entry individually, requiring $O(N)$ calls to the function $f$. Grover's algorithm, however, solves this problem using just $O(\sqrt{N})$ evaluations, offering a quadratic improvement over classical methods.

Grover's algorithm starts by preparing an equal superposition of all possible states in the search space, assuming $N=2^n$:
\begin{equation}
    \ket{s} = \frac{1}{\sqrt{N}}\sum_{i=0}^{N-1}\ket{i}
\end{equation}
It is convenient to express $\ket{s}$ in terms of two orthogonal basis vectors: the target state $\ket{x}$ satisfying $f(x)=1$, and $\ket{x^\perp}$, representing all states that do not satisfy the condition. Thus:
\begin{equation}
    \ket{s} = \frac{1}{\sqrt{N}}\ket{x} + \sqrt{\frac{N-1}{N}}\ket{x^\perp}
\end{equation}
Measuring this initial superposition directly yields the target state $\ket{x}$ with probability $1/N$, no better than a classical random guess. We then define two quantum operators:

The phase oracle $U_f$:
    \begin{equation}
    U_f \ket{i} = (-1)^{f(i)}\ket{i}
\end{equation}
which applies a phase shift of $-1$ exclusively to the target state $\ket{x}$.

And the diffusion operator $D$:
\begin{equation}
    D = H^{\otimes n}U_sH^{\otimes n}
\end{equation}
where $H^{\otimes n}$ denotes the Hadamard transformation over each qubit, and $U_s$ applies a negative ($-1$) phase shift exclusively to the state $\ket{0}$. The operator $D$ reflects each state's amplitude about the mean amplitude of the entire superposition. Consequently, states with negative phases (such as the marked state) have their amplitudes amplified, while those with positive phases are reduced.

These two operators combine into a single Grover iteration $G$:
\begin{equation}
    G = DU_f
\end{equation}
Repeated application of $G$ incrementally amplifies the amplitude of $\ket{x}$ and diminishes that of $\ket{x^\perp}$. After $t$ iterations, the state evolves to:
\begin{equation}
    G^t\ket{s} = sin((2t+1)\theta)\ket{x} + cos((2t+1)\theta)\ket{x^\perp}
\end{equation}
where $\theta = sin^{-1}(1/\sqrt{N})$.

The optimal iteration count $t$, maximizing the probability of successfully measuring $\ket{x}$, is found by solving:
\begin{equation}
    sin((2t+1)\theta) = 1 \;\;\Rightarrow \;\; (2t+1)\theta = \frac{\pi}{2}
\end{equation}
Solving for $t$ gives:
\begin{equation}
    t = \lfloor\frac{\pi}{4\theta} - \frac{1}{2}\rfloor
\end{equation}
Since $t$ must be an integer, rounding introduces a slight error, making Grover’s algorithm inherently probabilistic rather than fully deterministic.

However, an interesting special case arises when the search space size $N$ equals $4$, or equivalently, when exactly $1/4$ of the states satisfy the condition $f(i)=1$. In this scenario $\theta = sin^{-1}(1/2) = \pi/6$, and the optimal iteration count simplifies exactly to:
\begin{equation}
    t = \frac{6\pi}{4\pi} - \frac{1}{2} = 1
\end{equation}
Here, after precisely one iteration, the target state amplitude reaches unity, resulting in deterministic success. In fact, this is the only non-trivial case in which the original algorithm is exact \cite{groverExact}. This unique property forms the conceptual basis for our proposed deterministic quantum search algorithm, in which we iteratively mark $1/4$ of the search space, enabling its deterministic reduction.

The general concept of achieving deterministic quantum search by progressively marking and reducing one-quarter of the search space was previously proposed in~\cite{exponentialQuantumSearch}. That approach introduced an augmented oracle that marked the target state along with the first $N/4 - 1$ states in the superposition, followed by a specialised diffusion operator reflecting only around the currently relevant subspace. Subsequently,~\cite{exponentialRebuttle} showed that constructing such a selective diffusion operator would require $O(N^{\log_2(3)/2})$ queries to the base oracle, matching the query complexity of our approach.

Despite this similarity in scaling, our approach differs fundamentally. Rather than marking the first $N/4-1$ states, which also risks failure if the target state $x$ lies among them, we construct an augmented oracle that marks all states sharing the first two bits with $x$, corresponding precisely to one-quarter of the search space. This strategy enables deterministic reduction of the superposition using only the standard 2-qubit diffusion operator, avoiding the need for selective global reflections, which are extremely costly to implement in practice. Moreover, it naturally solves the partial quantum search problem as an interim step, further enhancing both the utility and applicability of our method.

\section{Deterministic Quantum Search}
In this section, we present our deterministic quantum search algorithm. We show how to recursively construct an oracle that marks exactly one-quarter of the states in the search space, enabling iterative and deterministic reduction until only the target state remains.

We first define a generalised prefix-matching function $g(x, m)$:
\begin{align}
    g(i, m) &= \begin{cases}
    1, & \text{if the first } n-m \text{ bits of } i \text{ and } x \text{ match }, \\
    0, & \text{otherwise}
    \end{cases}
\end{align}
Note that $f(i)=g(i,0)$.

The corresponding oracle $U_m$, which applies a negative phase to all states matching the prefix, is defined as:
\begin{equation}
U_m\ket{i} = (-1)^{g(i,m)}\ket{i}, \quad U_0=U_f.
\end{equation}

We also define the 2-qubit diffusion operator $D_2$:
\begin{equation}
    D_2 = H^{\otimes 2}U_sH^{\otimes 2}
\end{equation}
where $U_s$ applies a  phase exclusively to the state $\ket{00}$.

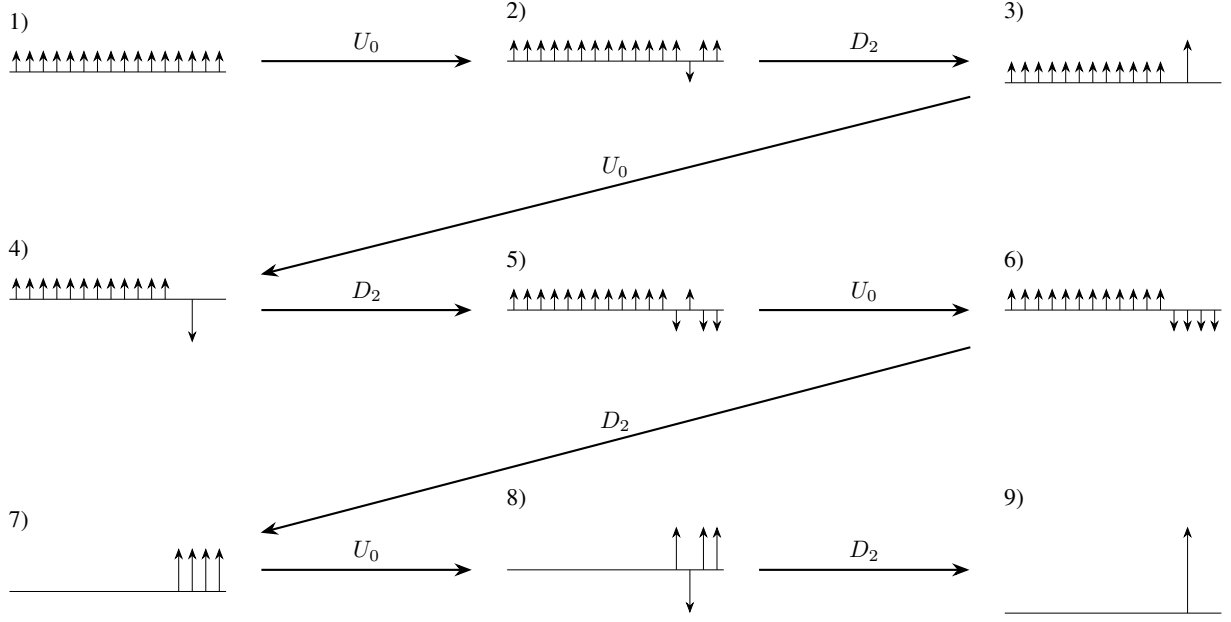
\begin{figure}

\begin{tikzpicture}[>=Stealth, node distance=2.5cm and 3.5cm, every node/.style={scale=0.9}]
  
  \newcommand{\arrowstate}[1]{
    \begin{tikzpicture}[scale=0.4]
      \draw[solid] (-4,0) -- (4,0); 
      #1
    \end{tikzpicture}
  }


  \node (n1) at (0, 0) {\arrowstate{
    \foreach \i in {0,...,15} {
      \pgfmathsetmacro{\x}{-3.75 + \i * 0.5}
      \draw[->] (\x,0) -- ++(0,0.8);
    }
  }};

  \node (n2) [right=of n1] {\arrowstate{
    \foreach \i/\h in {
      0/0.8, 1/0.8, 2/0.8, 3/0.8, 4/0.8, 5/0.8, 6/0.8, 7/0.8,
      8/0.8, 9/0.8, 10/0.8, 11/0.8, 12/0.8, 13/-0.8, 14/0.8, 15/0.8
    } {
      \pgfmathsetmacro{\x}{-3.75 + \i * 0.5}
      \draw[->] (\x,0) -- ++(0,\h);
    }
  }};

  \node (n3) [right=of n2] {\arrowstate{
    \foreach \i/\h in {
      0/0.8, 1/0.8, 2/0.8, 3/0.8, 4/0.8, 5/0.8, 6/0.8, 7/0.8,
      8/0.8, 9/0.8, 10/0.8, 11/0.8, 12/0.0, 13/1.6, 14/0.0, 15/0.0
    } {
      \pgfmathsetmacro{\x}{-3.75 + \i * 0.5}
      \ifdim \h pt = 0pt
      \else
        \draw[->] (\x,0) -- ++(0,\h);
      \fi
    }
  }};

  \node (n4) [below=2.5cm of n1] {\arrowstate{
    \foreach \i/\h in {
      0/0.8, 1/0.8, 2/0.8, 3/0.8, 4/0.8, 5/0.8, 6/0.8, 7/0.8,
      8/0.8, 9/0.8, 10/0.8, 11/0.8, 12/0.0, 13/-1.6, 14/0.0, 15/0.0
    } {
       \pgfmathsetmacro{\x}{-3.75 + \i * 0.5}
      \ifdim \h pt = 0pt
      \else
        \draw[->] (\x,0) -- ++(0,\h);
      \fi
    }
  }};

  \node (n5) [right=of n4] {\arrowstate{
    \foreach \i/\h in {
      0/0.8, 1/0.8, 2/0.8, 3/0.8, 4/0.8, 5/0.8, 6/0.8, 7/0.8,
     8/0.8, 9/0.8, 10/0.8, 11/0.8, 12/-0.8, 13/0.8, 14/-0.8, 15/-0.8
    } {
      \pgfmathsetmacro{\x}{-3.75 + \i * 0.5}
      \draw[->] (\x,0) -- ++(0,\h);
    }
  }};

  \node (n6) [right=of n5] {\arrowstate{
    \foreach \i/\h in {
      0/0.8, 1/0.8, 2/0.8, 3/0.8, 4/0.8, 5/0.8, 6/0.8, 7/0.8,
     8/0.8, 9/0.8, 10/0.8, 11/0.8, 12/-0.8, 13/-0.8, 14/-0.8, 15/-0.8
    } {
      \pgfmathsetmacro{\x}{-3.75 + \i * 0.5}
      \ifdim \h pt = 0pt
      \else
        \draw[->] (\x,0) -- ++(0,\h);
      \fi
    }
  }};

  \node (n7) [below=2.5cm of n4] {\arrowstate{
    \foreach \i/\h in {
      0/0.0, 1/0.0, 2/0.0, 3/0.0, 4/0.0, 5/0.0, 6/0.0, 7/0.0,
      8/0.0, 9/0.0, 10/0.0, 11/0.0, 12/1.6, 13/1.6, 14/1.6, 15/1.6
    } {
       \pgfmathsetmacro{\x}{-3.75 + \i * 0.5}
      \ifdim \h pt = 0pt
      \else
        \draw[->] (\x,0) -- ++(0,\h);
      \fi
    }
  }};

    \node (n8) [right=of n7] {\arrowstate{
    \foreach \i/\h in {
      0/0.0, 1/0.0, 2/0.0, 3/0.0, 4/0.0, 5/0.0, 6/0.0, 7/0.0,
      8/0.0, 9/0.0, 10/0.0, 11/0.0, 12/1.6, 13/-1.6, 14/1.6, 15/1.6
    } {
       \pgfmathsetmacro{\x}{-3.75 + \i * 0.5}
      \ifdim \h pt = 0pt
      \else
        \draw[->] (\x,0) -- ++(0,\h);
      \fi
    }
  }};

      \node (n9) [right=of n8] {\arrowstate{
    \foreach \i/\h in {
      0/0.0, 1/0.0, 2/0.0, 3/0.0, 4/0.0, 5/0.0, 6/0.0, 7/0.0,
      8/0.0, 9/0.0, 10/0.0, 11/0.0, 12/0.0, 13/3.2, 14/0.0, 15/0.0
    } {
       \pgfmathsetmacro{\x}{-3.75 + \i * 0.5}
      \ifdim \h pt = 0pt
      \else
        \draw[->] (\x,0) -- ++(0,\h);
      \fi
    }
  }};

\draw[->, thick, shorten >=10pt, shorten <=10pt] (n1) -- node[above] {$U_0$} (n2);

\draw[->, thick, shorten >=10pt, shorten <=10pt] (n2) -- node[above] {$D_2$} (n3);

\draw[->, thick, shorten >=10pt, shorten <=10pt] (n3) -- node[above] {$U_0$} (n4);

\draw[->, thick, shorten >=10pt, shorten <=10pt] (n4) -- node[above] {$D_2$} (n5);

\draw[->, thick, shorten >=10pt, shorten <=10pt] (n5) -- node[above] {$U_0$} (n6);

\draw[->, thick, shorten >=10pt, shorten <=10pt] (n6) -- node[above] {$D_2$} (n7);

\draw[->, thick, shorten >=10pt, shorten <=10pt] (n7) -- node[above] {$U_0$} (n8);

\draw[->, thick, shorten >=10pt, shorten <=10pt] (n8) -- node[above] {$D_2$} (n9);

\node[anchor=south west] at (n1.north west) { 1)};

\node[anchor=south west] at (n2.north west) { 2)};

\node[anchor=south west] at (n3.north west) {  3)};

\node[anchor=south west] at (n4.north west) {  4)};

\node[anchor=south west] at (n5.north west) {  5)};

\node[anchor=south west] at (n6.north west) {  6)};

\node[anchor=south west] at (n7.north west) {  7)};

\node[anchor=south west] at (n8.north west) {  8)};

\node[anchor=south west] at (n9.north west) {  9)};

\end{tikzpicture}

\caption{Step-by-step evolution of state amplitudes throughout the search process. Step 1 starts with an equal superposition over all states. Steps 2–6 show the construction of the oracle $U_2$ using three applications of $U_0$ and two of $D_2$. Steps 7–9 demonstrate how the resulting oracle can be used to collapse the superposition onto the target state with certainty.}

    \label{fig:searchProcess}
\end{figure}

Our goal is to recursively build the oracle $U_{n-2}$, marking exactly $1/4$ of the states in the search space (those sharing the first two bits with $x$). We illustrate the method by first constructing $U_2$ from $U_0$ ($U_f$).

Start with an $n = log_2(N)$ qubit uniform superposition:
\begin{equation}
\ket{s} = \frac{1}{\sqrt{2^n}}\sum_{i=0}^{2^n-1}\ket{i}.
\end{equation}

Apply $U_0$ to mark the target state $\ket{x}$:
\begin{equation}
U_0\ket{s} = \frac{1}{\sqrt{2^n}}\sum_{i=0}^{2^n-1}(-1)^{g(i, 0)}\ket{i}.
\end{equation}

For clarity, we can rewrite the state in the terms of two subsystems $\ket{i}$ (first $n-2$ qubits) and $\ket{j}$ (last 2 qubits):
\begin{equation}
U_0\ket{s} = \frac{1}{\sqrt{2^n}}\sum_{i=0}^{2^{n-2}-1}\sum_{j=0}^{2^2-1} \ket{i} \otimes (-1)^{g(i\|j, 0)}\ket{j}.
\end{equation}
where $i \| j$ denotes the bit string concatenation of $i$ and $j$.

Apply $D_2$ to $\ket{j}$ to amplify the state that satisfies $g(i, 0)$ ($\ket{x}$) and reduce the amplitude of the other states that satisfy $g(i, 2)$ to 0:
\begin{equation}
    (I^{\otimes n-2} \otimes D_2) U_0 \ket{s} 
    = \frac{-1}{\sqrt{2^n}} \sum_{i=0}^{2^{n-2}-1} \sum_{j=0}^{2^2-1} \ket{i} \otimes \left[\ket{j} - g(i,2)\ket{j} + 2g(i\|j,0)\ket{j}\right].
\end{equation}

Apply $U_0$ again to invert the phase of $\ket{x}$:
\begin{equation}
    U_0(I^{\otimes n-2} \otimes D_2) U_0 \ket{s} 
    = \frac{-1}{\sqrt{2^n}} \sum_{i=0}^{2^{n-2}-1} \sum_{j=0}^{2^2-1} \ket{i} \otimes \left[\ket{j} - g(i,2)\ket{j} - 2g(i\|j,0)\ket{j}\right].
\end{equation}

A second application of $D_2$ reverses the initial amplification, but with the phase on states neighbouring $\ket{x}$ flipped:
\begin{equation} (I^{\otimes n-2} \otimes D_2)U_0(I^{\otimes n-2} \otimes D_2)U_0\ket{s} = \frac{1}{\sqrt{2^n}}\sum_{i=0}^{2^{n-2}-1}\sum_{j=0}^{2^2-1}\ket{i} \otimes (-1)^{g(i,2) \oplus g(i\|j,0)}\ket{j}  \end{equation}
where $\oplus$ is the boolean XOR operation.

A final $U_0$ aligns the phase of $\ket{x}$:
\begin{equation}
\begin{split}
U_0(I^{\otimes n-2} \otimes D_2)U_0(I^{\otimes n-2} \otimes D_2)U_0\ket{s} &= \frac{1}{\sqrt{2^n}}\sum_{i=0}^{2^{n-2}-1}\sum_{j=0}^{2^2-1}\ket{i} \otimes (-1)^{g(i,2)}\ket{j}\\
&= \frac{1}{\sqrt{2^n}}\sum_{i=0}^{2^{n-2}-1}\sum_{j=0}^{2^2-1}(-1)^{g(i,2)}\ket{i} \otimes \ket{j}\\
&= U_2 \ket{s}
\end{split}
\end{equation}
Thus, using only three applications of the initial oracle $U_0$, we successfully construct $U_2$, which marks all states whose first $n-2$ bits match those of the target state $x$. This construction process is illustrated in Steps 1–6 of Figure~\ref{fig:searchProcess}.

Given the oracle operator $U_m$, the operator $U_{m+2}$ can be constructed with three applications of $U_m$ and two of $D_2$.

We can again rewrite the initial superposition $\ket{s}$, this time in terms of three subsystems $\ket{i}$ (first $n-(m+2)$ qubits), $\ket{j}$ (the next 2 qubits) and $\ket{k}$ (the final $m$ qubits):
\begin{equation}
\ket{s} = \frac{1}{\sqrt{2^n}}\sum_{i=0}^{2^{n-m-2}-1}\sum_{j=0}^{2^2-1}\sum_{k=0}^{2^{m}-1} \ket{i} \otimes \ket{j} \otimes \ket{k}
\end{equation}
Apply $U_m$ to mark all states prefixed by the same $n-m$ bits as $x$:
\begin{equation}
\ket{s} = \frac{1}{\sqrt{2^n}}\sum_{i=0}^{2^{n-m-2}-1}\sum_{j=0}^{2^2-1}\sum_{k=0}^{2^{m}-1} \ket{i} \otimes (-1^{g(i\|j,m)}) \ket{j} \otimes \ket{k}
\end{equation}
Apply $D_2$ to $\ket{j}$ to amplify the states that satisfy $g(i, m)$ and reduce the amplitude of the other states that satisfy $g(i, m+2)$ to 0:
\begin{equation}
\begin{aligned}
    &(I^{\otimes n-m-2} \otimes D_2 \otimes I^{\otimes m})U_m\ket{s} \\
    &\quad \begin{aligned}
        &= \frac{-1}{\sqrt{2^n}}\sum_{i=0}^{2^{n-m-2}-1}\sum_{j=0}^{2^{2}-1}\sum_{k=0}^{2^{m}-1}\ket{i} \otimes (\ket{j} - g(i,m+2)\ket{j} + 2g(i\|j,m)\ket{j}) \otimes \ket{k}
    \end{aligned}
\end{aligned}
\end{equation}
Apply $U_m$ again to invert the phase of the states satisfying $g(i\|j,m)$:
\begin{equation}
\begin{aligned}
    &U_m(I^{\otimes n-m-2} \otimes D_2 \otimes I^{\otimes m})U_m\ket{s} \\
    &\quad \begin{aligned}
        &= \frac{-1}{\sqrt{2^n}}\sum_{i=0}^{2^{n-m-2}-1}\sum_{j=0}^{2^{2}-1}\sum_{k=0}^{2^{m}-1}\ket{i} \otimes (\ket{j} - g(i,m+2)\ket{j} - 2g(i\|j,m)\ket{j}) \otimes \ket{k}
    \end{aligned}
\end{aligned}
\end{equation}
As before, a second application of $D_2$ reverses the original amplification, but with the phase on the affected states flipped:
\begin{equation}
\begin{aligned}
    &(I^{\otimes n-m-2} \otimes D_2 \otimes I^{\otimes m})U_m(I^{\otimes n-m-2} \otimes D_2 \otimes I^{\otimes m})U_m\ket{s} \\
    &\quad \begin{aligned}
        &= \frac{1}{\sqrt{2^n}}\sum_{i=0}^{2^{n-m-2}-1}\sum_{j=0}^{2^{2}-1}\sum_{k=0}^{2^{m}-1}\ket{i} \otimes (-1^{g(i,m+2) \oplus g(i\|j, m)})\ket{j} \otimes \ket{k}
    \end{aligned}
\end{aligned}
\end{equation}
Apply $U_m$ one more time to align the phases:
\begin{equation}
\begin{aligned}
    &U_m(I^{\otimes n-m-2} \otimes D_2 \otimes I^{\otimes m})U_m(I^{\otimes n-m-2} \otimes D_2 \otimes I^{\otimes m})U_m\ket{s} \\
    &\quad \begin{aligned}
        &= \frac{1}{\sqrt{2^n}}\sum_{i=0}^{2^{n-m-2}-1}\sum_{j=0}^{2^{2}-1}\sum_{k=0}^{2^{m}-1}\ket{i} \otimes (-1^{g(i,m+2)})\ket{j} \otimes \ket{k} \\
        &= \frac{1}{\sqrt{2^n}}\sum_{i=0}^{2^{n-m-2}-1}\sum_{j=0}^{2^{2}-1}\sum_{k=0}^{2^{m}-1}(-1^{g(i,m+2)})\ket{i} \otimes \ket{j} \otimes \ket{k}\\
        &= U_{m+2}\ket{s}
    \end{aligned}
\end{aligned}
\end{equation}
Thus, the oracle $U_{m+2}$ is successfully constructed using three applications of $U_m$ and two of $D_2$.

By recursion, the general oracle $U_m$ can be built entirely from the base oracle $U_0$ and the two-qubit diffusion operator $D_2$, as defined by:
\begin{equation}
U_m = U_{m-2}(I^{\otimes n-m-2}\otimes D_2 \otimes I^{\otimes m})U_{m-2}(I^{\otimes n-m-2}\otimes D_2 \otimes I^{\otimes m})U_{m-2}.
\label{eq:oracle_equation}
\end{equation}

With the oracle construction defined, we now apply it as the first step in the deterministic search process. We begin by constructing and applying the oracle operator $U_{n-2}$. This operator marks precisely those states whose first two bits match those of the target state $x$, encompassing exactly $1/4$ of the search space. Subsequently, applying the two-qubit diffusion operator $D_2$ to the two most significant qubits amplifies the amplitudes of these marked states while reducing all others to zero. Thus, the two most significant qubits become pure computational basis states corresponding to the first two bits of $x$:
\begin{equation}
    (D_2 \otimes I^{\otimes n-2})U_{n-2}\ket{s} = \ket{x_0x_1} \otimes \frac{1}{\sqrt{2^{n-2}}}\sum_{i=0}^{2^{n-2}-1}\ket{i}
\end{equation}
Here, $\ket{x_0x_1}$ explicitly denotes the computational basis state defined by the two most significant bits of $x$.

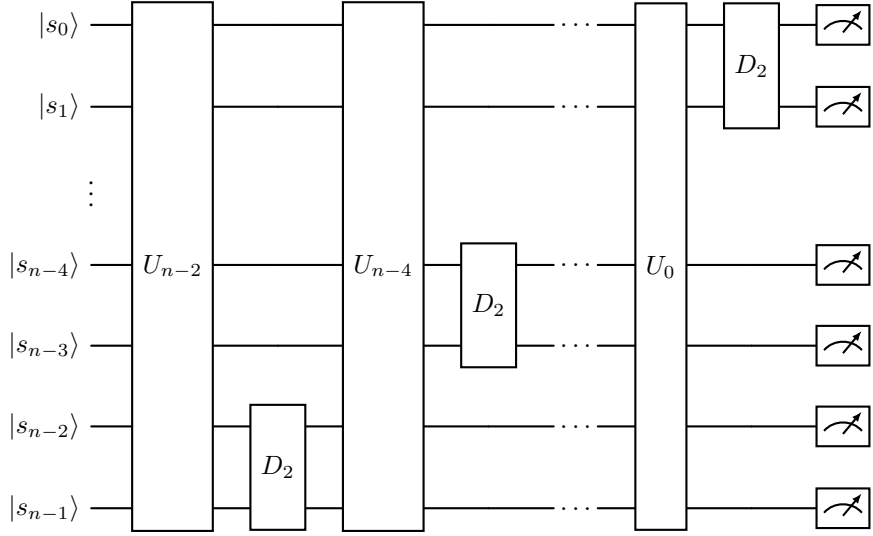
\begin{figure}[t]
\centering
\begin{quantikz}
\lstick{$\ket{s_0}$}&\gate[7]{U_{n-2}}&&\gate[7]{U_{n-4}}& & \ \ldots\ &\gate[7]{U_{0}}&\gate[2]{D_2}&\meter{}\\
\lstick{$\ket{s_1}$}&&&&& \ \ldots\ &&&\meter{}\\
\vdots\\
\lstick{$\ket{s_{n-4}}$}&&&&\gate[2]{D_2}& \ \ldots\ &&&\meter{}\\
\lstick{$\ket{s_{n-3}}$}&&&&& \ \ldots\ &&&\meter{}\\
\lstick{$\ket{s_{n-2}}$}&&\gate[2]{D_2}&&& \ \ldots\ &&&\meter{}\\
\lstick{$\ket{s_{n-1}}$}&&&&& \ \ldots\ &&&\meter{}
\end{quantikz}
\caption{\label{fig:search_circuit} Quantum circuit for deterministically finding a single element $x$ marked by the oracle $U_0$. The pattern of alternating $U_{m}$ and $D_2$ operations continues down to $U_0$, reducing the search space by one-quarter at each step. Final measurement yields the target state $\ket{x}$ with certainty.}

\end{figure}

The next two significant qubits can similarly be isolated by applying the oracle operator $U_{n-4}$, followed by the diffusion operator $D_2$ acting on them. By repeatedly applying this procedure, specifically, iterating the oracle $U_{n-2-2i}$ and the corresponding diffusion operator $(I^{\otimes 2i} \otimes D_2 \otimes I^{\otimes n-2i-2})$, we iteratively reduce the search space by a factor of $1/4$ at each step, until only the target state $\ket{x}$ remains. Formally, this iterative procedure is expressed as:
\begin{equation}
    [\prod_{i=0}^{n/2-1}(I^{\otimes 2i} \otimes D_2 \otimes I^{\otimes n-2i-2})U_{n-2i-2}]\ket{s} = \ket{x}.
    \label{eq:search_equation}
\end{equation}
This construction is illustrated in Figure~\ref{fig:search_circuit}.

Through this procedure, the target state $\ket{x}$ is recovered deterministically, overcoming the probabilistic limitation of Grover's original algorithm. By recursively constructing oracles that mark exactly one-quarter of the search space and applying structured, local diffusion steps, the algorithm guarantees success without requiring arbitrary phase rotations, relying only on the standard oracle and two-qubit diffusion operator.

\section{Analysis and Discussion}

We now analyse the complexity of our algorithm, focusing on the number of oracle calls to $U_0$ and applications of the two-qubit diffusion operator $D_2$. We then show how the method can be naturally adapted for partial database search, maintaining deterministic success. Finally, we discuss the practicality of the algorithm, comparing the implementation complexity of our approach to Grovers, and outline directions for future refinement and research.

\subsection{Algorithm Complexity}

Our algorithm constructs the oracle $U_{n-2}$ recursively, starting from the base oracle $U_0$. At each level of recursion, an oracle $U_m$ is built using three calls to $U_{m-2}$ (Eq.~\ref{eq:search_equation}). To construct $U_{n-2}$, the total number of $U_0$ calls is then $3^{((n-2)/2)}$. Since the algorithm applies increasingly smaller instances of $U_m$, the total number of oracle calls across all levels is:
\begin{equation}
    \sum_{i=1}^{n/2-1}3^{i} = \frac{3^{n/2}-3}{2} = O(\sqrt{3}^n) = O(N^{\log_2 (3) / 2}) \approx O(N^{0.7925}),
\end{equation}
where $N = 2^n$ is the size of the search space.

The number of two-qubit diffusion operators $D_2$ can be calculated similarly. Both the expanded oracle $U_m$ and the overall search algorithm constructions alternate between applying the base oracle and diffusion operators. While each expanded oracle involves one more $U_0$ call than $D_2$, the alternating pattern between $U_m$ and $D_2$ during the search procedure results in the total number of oracle $U_0$ calls and diffusion operator $D_2$ applications being equal.

Combining both analyses, our deterministic algorithm requires $O(N^{\log_2 (3) / 2}) \approx O(N^{0.7925})$ calls to the base oracle $U_0$ and an equal number of applications of the two-qubit diffusion operator $D_2$.

\subsection{Partial Database Search}
The recursive oracle construction naturally extends to the problem of \textit{partial database search}, where the goal is to identify only the first $k$ bits of the target item $x$~\cite{partialSearchDefinition}. Rather than locating the exact position of the target, this problem involves identifying the broader section or block of the database in which it resides. Such scenarios are common in tasks like spatial searches, decision trees, and classification, where only a subset of the target information is required. 

In our algorithm, each iteration of Equation~\ref{eq:search_equation} deterministically reveals two additional bits of the target item. After $i$ iterations, the state is projected onto the subspace where all elements share the first $k=2i$ bits with $x$. By terminating the process early, the algorithm performs a partial search, offering flexible control over the resolution of the target value. 

If the desired prefix length $k$ is not divisible by two, indexing can be offset by one bit by introducing an ancilla qubit and conditioning the initial oracle $U_0$ on it, allowing for arbitrary prefix lengths.

A key feature of this approach is the uniform superposition it generates over all states within the target block. In contrast, other partial search algorithms~\cite{partialSearchDefinition, partialSearch, LocalDeterminism2, partialSearchDeterministic} typically yield amplitude distributions biased toward the target. A uniform distribution is especially useful in later computations that need unbiased sampling or equal weighting of outcomes within the target block. While similar uniformity could be recovered by measuring and reinitialising the quantum state, this would add additional potential sources of error and may not be feasible in all contexts.

While this method requires fewer oracle calls than a full search, it still has a query complexity of $O(N^{log_2(3)/2})$. Alternative partial search methods can achieve lower query complexity, $O(\sqrt{N})$, comparable to traditional Grover’s algorithm. However, these approaches typically guarantee deterministic success only for specific distributions where the proportion of target states is exactly $1/116$ \cite{LocalDeterminism2}, or they rely on precise phase rotations \cite{partialSearchDeterministic}. In contrast, our recursive approach ensures deterministic success without such constraints, making it simpler and more practical to implement on quantum hardware.

\subsection{Implementation Complexity}
While Grover’s algorithm has a lower asymptotic query complexity ($O(N^{0.5})$ vs.\ $O(N^{0.7925})$ for our approach), its global diffusion operator must be decomposed into sequences of one- and two-qubit gates for execution on real hardware, introducing substantial overhead to the algorithm’s total resource cost. Specifically, implementing an $n$-controlled operation requires $O(n)$ two-qubit gates when ancillary qubits are permitted \cite{mcx1, mcx2}, and $O(n^2)$ in the absence of ancillae \cite{mcx3}. Since two-qubit gates are typically the most error-prone and difficult-to-implement operations on current quantum devices, they are a key determinant of a quantum algorithm's practicality. This issue is further exacerbated on hardware architectures with limited qubit connectivity, where controlled operations between distant qubits require multiple SWAP gates to facilitate. Our aim was to assess whether our algorithm’s use of two-qubit nearest-neighbour diffusion operations could yield shallower circuits and lower two-qubit gate counts after realistic decompositions, despite the higher theoretical query complexity.

To evaluate this, we compared the number of two-qubit gates required by fully decomposed implementations of our recursive quantum search algorithm and Grover’s algorithm. Simulations were performed using IBM’s Qiskit framework. Both algorithms were transpiled onto two representative hardware topologies: a grid coupling map inspired by the IBM \textit{ibm\_marrakesh} device~\cite{ibmq_marrakesh}, and an idealised fully connected topology. Identical gate bases were used in all cases, with CZ as the native two-qubit operation. Two scenarios were analysed: one including a toy oracle marking the $\ket{0}$ state, and another evaluating the diffusion operators alone, where the oracle was replaced with a barrier to preserve sequence order without adding cost.

\begin{figure}[t]
\centering
\includegraphics[width=1\textwidth]{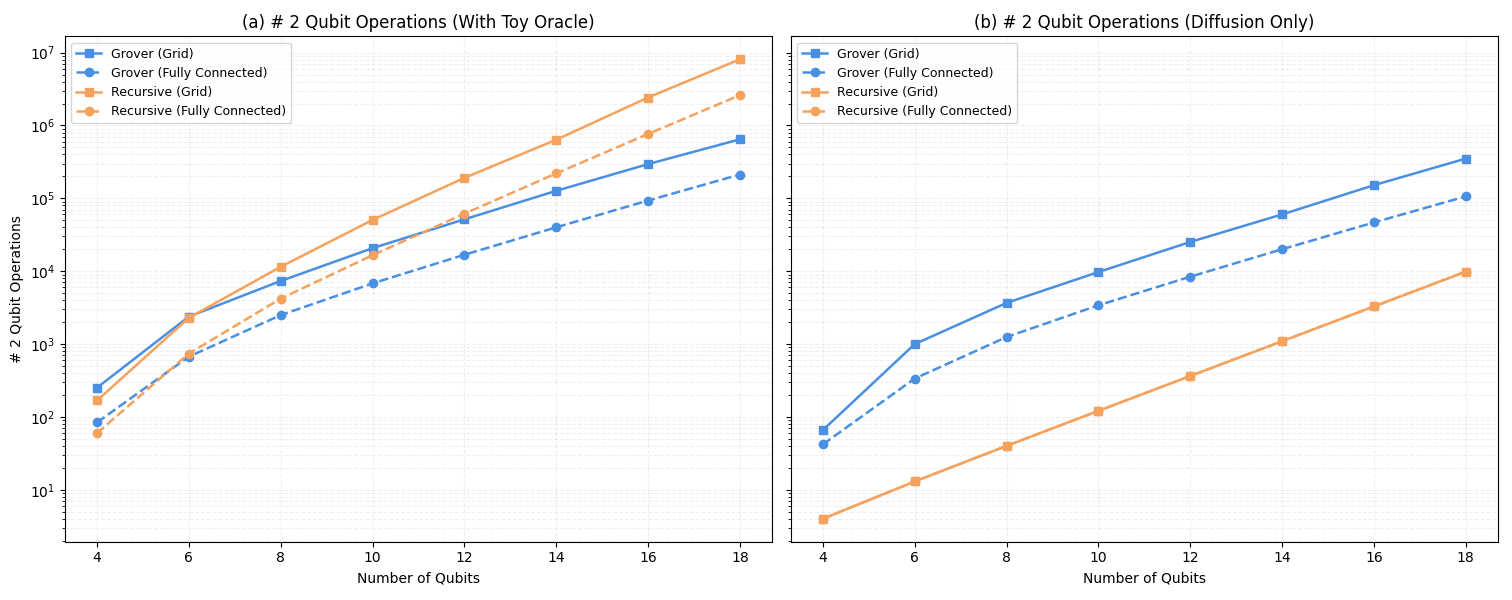}
\caption{Scaling of two-qubit gate counts for Grover’s algorithm and the proposed recursive algorithm, plotted on a logarithmic scale as a function of the number of qubits. Solid and dashed curves indicate performance on a grid and fully connected topology, respectively. (a) includes both diffusion and oracle complexity; (b) evaluates the diffusion steps alone. The diffusion steps in the recursive approach require significantly fewer operations to implement due to local, nearest-neighbour constructions; however, total two-qubit gate counts increase rapidly when oracle cost is included, underscoring the critical role of oracle implementation in practical performance.}
\label{fig:depth-scaling}
\end{figure}

Figure~\ref{fig:depth-scaling} presents the scaling of two-qubit gate counts for both algorithms with increasing numbers of qubits.

Figure~\ref{fig:depth-scaling} (a) shows the total two-qubit gate counts for fully decomposed circuits including the toy oracle. In the smallest test case of 4 qubits, our recursive scheme uses fewer two-qubit gates than Grover’s algorithm across both topologies. However, beyond 6 qubits, the number of two-qubit gates in our algorithm grows quickly, eventually surpassing Grover’s, highlighting the higher query complexity's dominant contribution to overall resource requirements. As expected, grid coupling results in greater two-qubit gate counts than the fully connected topology, primarily due to additional SWAPs needed to implement non-neighbouring controlled operations. This illustrates the additional overhead and difficulties in efficiently implementing quantum search on hardware architectures with low qubit connectivity

Figure~\ref{fig:depth-scaling} (b) shows the two-qubit gate counts for the diffusion operators alone. Across all tested qubit sizes (4–18 qubits), our recursive diffusions require over an order of magnitude fewer two-qubit gates than Grover’s global diffusion, demonstrating the clear advantage of local, nearest-neighbour diffusion. Furthermore, because the recursive diffusions rely exclusively on interactions between neighbouring qubits, their cost remains identical across grid and fully connected topologies, confirming their suitability for efficient hardware implementation across different architectures, including those with low qubit connectivity. In contrast, Grover’s global diffusion incurs a significantly higher gate cost on grid coupling due to the need for additional SWAPs to facilitate long-range controlled operations.

These results underscore a key trade-off: while our recursive diffusions enable much shallower implementations of the diffusion step than Grover’s, the total two-qubit gate count is ultimately dominated by the oracle cost. If constant-depth oracle operators can be realised, the benefits of shallow local diffusions could become decisive, particularly on near-term devices where minimising circuit depth is critical to counteract decoherence and gate errors. However, whether such efficient oracle constructions can be practically achieved remains an open question~\cite{quantumSearchPractical}.

Due to the asymptotically higher query complexity of our approach, the cost of the recursive diffusions will eventually exceed that of Grover’s at larger qubit counts. We can numerically estimate these boundaries for different multi-controlled gate decomposition: with $(n-2)$ ancillas and $6n - 6$ CX gates~\cite{mcx1}, the crossover occurs near $n=26$; with a single ancilla requiring $16n - 24$ CX gates~\cite{mcx2}, near $n=31$; and with a zero-ancilla $O(n^2)$ decomposition~\cite{mcx3}, beyond $n=36$. We note that these are approximations for fully connected hardware, for other topologies these boundaries would be greater.

\subsection{Discussion and Future Work}

Our deterministic quantum search algorithm addresses a fundamental limitation of quantum search by providing guaranteed success, thereby eliminating the inherent uncertainty of traditional approaches. This determinism is particularly valuable in sensitive applications such as cryptography and optimisation, where complete certainty is required.

A major advantage of our method is that it achieves determinism without resorting to arbitrary phase rotations required by other deterministic quantum search variants. Such arbitrary rotations typically demand high-fidelity, general-purpose quantum gates, posing considerable implementation challenges with current quantum hardware. Instead, our approach utilises exclusively standard local two-qubit diffusion operators, significantly simplifying gate decomposition and improving practical feasibility.

The algorithm's recursive oracle expansion strategy enhances its practical utility through modular and scalable construction. This design characteristic simplifies implementation and facilitates integration into broader quantum programs, directly addressing the key challenge of scaling precise operation sequences faced by previous deterministic quantum search methods~\cite{NearDeterminismLocal}.

Our performance evaluation underscores the benefits of using local nearest-neighbour diffusion operators. Across various qubit counts, our diffusion implementation requires over an order of magnitude fewer two-qubit gates than Grover's global diffusion. These findings align with prior studies highlighting the practical efficiency of local diffusion strategies~\cite{searchDepthOptimization, hardwareGrover2}. Moreover, since our diffusion operators rely exclusively on neighbouring-qubit interactions, their implementation costs remain stable across different hardware connectivity models. Conversely, global diffusion steps in Grover’s algorithm incur significant overhead on hardware with limited connectivity due to the additional SWAP operations necessary for implementing non-local interactions.

Nevertheless, our results also highlight an important consideration: despite the significant gate complexity reductions achieved in diffusion operations, the overall resource cost is ultimately dominated by oracle implementation due to our algorithm’s higher query complexity ($O(N^{0.7925})$. Therefore, fully leveraging the efficiency of local diffusion will depend critically on the development of low-overhead, constant-depth oracle implementations, a key open challenge for practical quantum search.

Additionally, extending our deterministic approach to accommodate multiple marked states represents a significant avenue for future research. If multiple targets are initially marked, the diffusion step may no longer preserve the recursive structure, impeding correct oracle expansion. A potential solution, assuming knowledge of the marked elements’ distribution, involves strategically reordering qubits to partition the search space into distinct blocks containing single marked items. Such a strategy would enable independent execution across blocks, potentially reducing complexity to approximately $O((\frac{N}{M})^{\log_2(3)/2})$, where $M$ is the number of marked states. However, full determinism would only be preserved if the number of blocks and marked elements is a power of two. Extending the algorithm to support unknown or non-uniform distributions of marked states remains a further interesting and open direction for future investigation. 

Finally, while we evaluate the two-qubit gate count and depth of our approach, a key next step is to evaluate the algorithm’s robustness under realistic noise models. Simulating or implementing the algorithm on near-term quantum hardware would provide valuable insight into its practical viability and error tolerance. Notably, the higher ceiling success probability may impart greater noise resilience, particularly for small-scale applications, warranting further investigation.  

Overall, our deterministic quantum search algorithm's unique combination of guaranteed outcomes, simplified implementation through local operations, and modular recursive structure provides substantial advantages for practical hardware implementation, laying a robust foundation for future advancements in quantum search approaches.

\section{Conclusion}

In this work, we have presented a novel deterministic quantum search algorithm that overcomes the inherent probabilistic limitations of traditional quantum search without requiring arbitrary phase rotations. Our approach uses a recursive construction based on local two-qubit diffusion operators to expand the base oracle so that it marks one-quarter of the search space at each step, enabling deterministic, step-by-step reduction of the superposition until the marked state can be measured with certainty.

The algorithm achieves a query complexity of $O(N^{\log_2(3)/2}) \approx O(N^{0.7925})$, representing a significant improvement over classical $O(N)$ scaling. Our simulations demonstrate that the local, nearest-neighbour diffusion steps in our algorithm require over an order of magnitude fewer two-qubit operations to implement than Grover’s global diffusion for search spaces up to at least 18 qubits, achieving even greater efficiencies on hardware architectures with limited connectivity. Combined with its fixed, low-depth construction and modular recursive design, this advantage makes our algorithm a promising candidate for practical implementation and a robust building block for integration into larger quantum algorithms.

We have further shown that the algorithm naturally supports partial database searches, broadening its applicability to scenarios where deterministic identification of selected target bits is required without performing a full search.

Looking ahead, it is proposed to investigate how to extend the algorithm to efficiently handle multiple target states, as well as explore its robustness under realistic noise models. Addressing these questions will further demonstrate the practical relevance of our approach and expand its potential impact on quantum search and related applications.

\bibliographystyle{unsrtnat}
\bibliography{search_paper/citations}

\end{document}